\documentclass[10pt, paper=a4, UKenglish]{article}
\usepackage{graphicx}

\def\Title#1{\begin{center} {\Large #1 } \end{center}}
\def\Author#1{\begin{center}{ \sc #1} \end{center}}
\def\Address#1{\begin{center}{ \it #1} \end{center}}

\newcommand\pubblock{\rightline{\begin{tabular}{l} Proceedings of the CTD 2022\\ \pubnumber\\
         \pubdate  \end{tabular}}}

\newenvironment{Abstract}{\begin{quotation} \begin{center} 
             \large ABSTRACT \end{center}\bigskip 
      \begin{center}\begin{large}}{\end{large}\end{center} \end{quotation}}

\newenvironment{Presented}{\begin{quotation} \begin{center} 
             PRESENTED AT\end{center}\bigskip 
      \begin{center}\begin{large}}{\end{large}\end{center} \end{quotation}}

\def\Acknowledgements{\bigskip  \bigskip \begin{center} \begin{large}
      \bf ACKNOWLEDGEMENTS \end{large}\end{center}}

\def\beq{\begin{equation}}
\def\eeq#1{\label{#1}\end{equation}}
\def\eeqn{\end{equation}}

\def\beqa{\begin{eqnarray}}
\def\eeqa#1{\label{#1}\end{eqnarray}}
\def\eeqan{\end{eqnarray}}

\let\bar=\overbar

\def\E{{\cal E}}

\def\Dslash{\not{\hbox{\kern-4pt $D$}}}
\def\dslash{\not{\hbox{\kern-2pt $\del$}}}

\def\msb{{\bar{\ssstyle M \kern -1pt S}}}

\usepackage{color}
\usepackage{lineno}
\usepackage{subfig}
\usepackage{amssymb}
\usepackage{hyperref}

\newcommand\pubnumber{PROC-CTD2022-32\\MIT-CTP/5476\\ DESY-22-113}

\newcommand\pubdate{\today}

\newcommand{\conference}{Connecting the Dots Workshop (CTD 2022)\\
May 31 - June 2, 2022}

\usepackage{fancyhdr}
\definecolor{mygrey}{RGB}{105,105,105}
\begin{document}

\large
\begin{titlepage}
\pubblock

\vfill
\Title{Track reconstruction at the LUXE experiment using quantum algorithms}
\vfill

\Author{Arianna Crippa$^{1,2}$, Lena Funcke$^3$, Tobias Hartung$^{4}$, Beate Heinemann$^{6,7}$, Karl Jansen$^1$, Annabel Kropf$^{6,7}$, Stefan K\"uhn$^5$, Federico Meloni$^6$, David Spataro$^{6,7}$, Cenk T\"uys\"uz$^{1,2}$  and Yee Chinn Yap$^6$}

\Address{$^1$ Deutsches Elektronen-Synchrotron DESY, Platanenallee 6, 15738 Zeuthen, Germany}
\Address{$^2$ Institut für Physik, Humboldt-Universit\"at zu Berlin, Newtonstr. 15, 12489 Berlin, Germany}
\Address{$^3$ Center for Theoretical Physics, Co-Design Center for Quantum Advantage, and NSF AI Institute for Artificial Intelligence and Fundamental Interactions, Massachusetts Institute of Technology, 77 Massachusetts Avenue, Cambridge, MA 02139, USA}
\Address{$^4$ Northeastern University - London, Devon House, St Katharine Docks, London, \mbox{E1W 1LP}, United Kingdom}
\Address{$^5$ Computation-Based Science and Technology Research Center, The Cyprus Institute, 20 Kavafi Street, 2121 Nicosia, Cyprus}
\Address{$^6$ Deutsches Elektronen-Synchrotron DESY, Notkestr. 85, 22607 Hamburg, Germany}
\Address{$^7$ Physikalisches Institut, Albert-Ludwigs-Universit\"at Freiburg, Hermann-Herder-Str. 3a, 79104 Freiburg, Germany}
\vfill
david.spataro@desy.de
\newpage

\begin{Abstract}
LUXE (Laser Und XFEL Experiment) is a proposed experiment at DESY which will study Quantum Electrodynamics (QED) in the strong-field regime, where QED becomes non-perturbative. Measuring the rate of created electron-positron pairs using a silicon pixel tracking detector is an essential ingredient to study this regime. Precision tracking of positrons traversing the four layers of the tracking detector becomes very challenging at high laser intensities due to the high rates, which can be computationally expensive for classical computers. In this work, we update our previous study of the potential of using quantum computing to reconstruct positron tracks. The reconstruction task is formulated as a quadratic unconstrained binary optimisation and is solved using simulated quantum computers and a hybrid quantum-classical algorithm, namely the variational quantum eigensolver. Different ansatz circuits and optimisers are studied. The results are discussed and compared with classical track reconstruction algorithms using a graph neural network and a combinatorial Kalman filter.
\end{Abstract}

\vfill

\begin{Presented}
\conference
\end{Presented}
\vfill
\end{titlepage}
\def\thefootnote{\fnsymbol{footnote}}
\setcounter{footnote}{0}

\normalsize

\section{Introduction}
\label{intro}

In this work, we present an update of our previous study of track reconstruction with quantum algorithms~\cite{ACAT} at LUXE.

LUXE~\cite{CDR} is a planned experiment at DESY in Hamburg to study the transition far into the strong-field regime of QED, where QED becomes non-perturbative. The classical non-linearity parameter,
\beq{  \xi = \frac{m_e}{\omega_L}\frac{\E_L}{\E_{cr}}}, 
\eeq{xi}
where $m_e$ is the electron mass, $\omega_L$ is the laser frequency, $\E_L$ is the instantaneous laser field strength, and $\E_{cr}$ = $m_e^2c^3 / e\hbar$ is the critical field strength, known as the Schwinger limit, is used to demarcate the regime of strong-field QED in particle-laser and photon-laser interactions ($\xi \gg 1$).

To study the phenomena occurring in these interactions, a key measurement is the number of positrons generated from the Breit-Wheeler process with respect to the parameter $\xi$. The number of expected positrons ranges from less than 10$^{-4}$ up to 10$^6$. Both a low background rate ($<$\ 10$^{-3}$) at low $\xi$ and a good linearity of the number of reconstructed tracks as a function of $\xi$ are essential. For this challenging task, we explore the potential of using quantum computing for track reconstruction. Other quantum computing algorithms studied for charged particle tracking can be found in Ref.~\cite{HGray} and references therein. 

\begin{figure}[!htb]
\centering
 \includegraphics[width=0.75\linewidth]{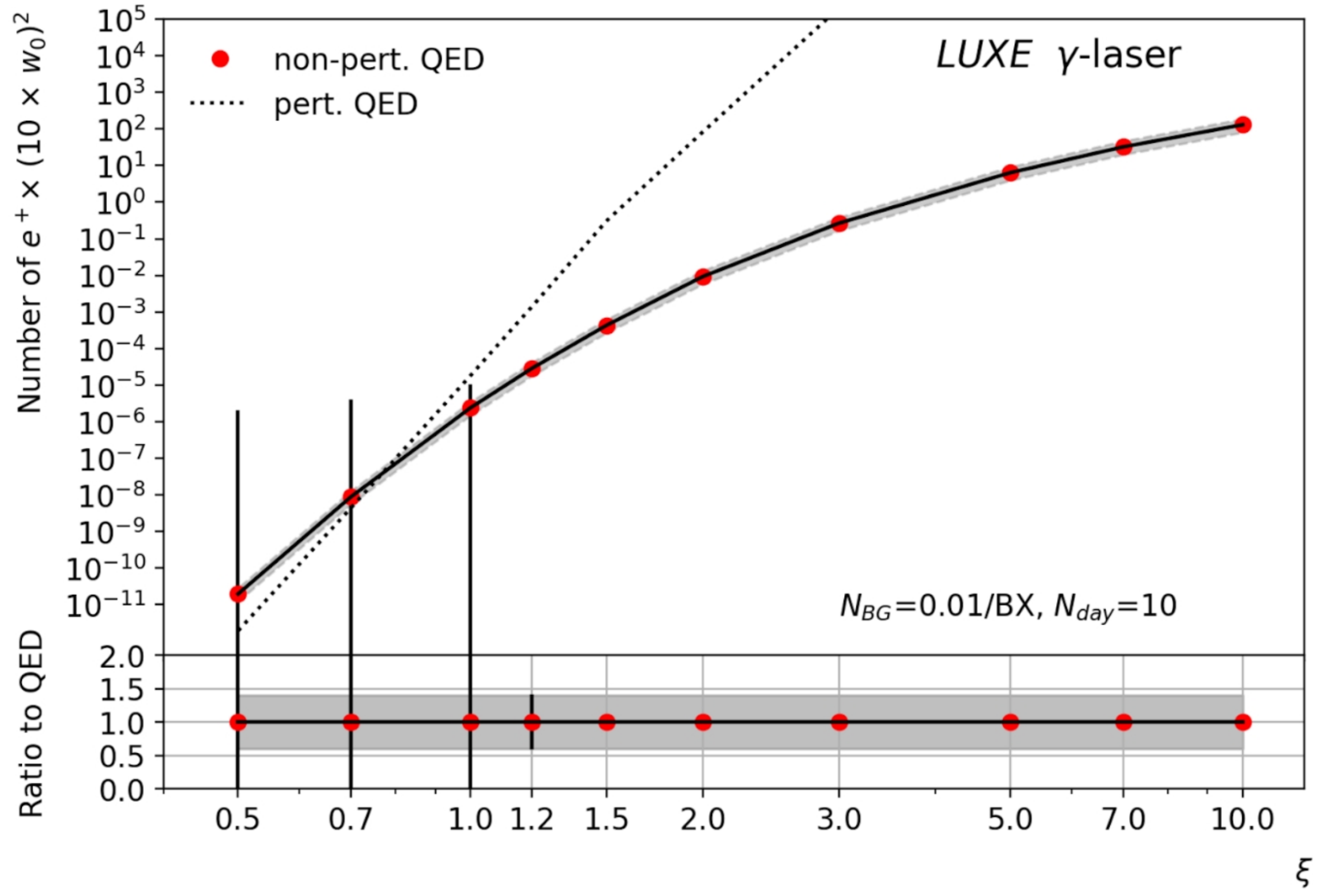}
 \caption{Number of positrons per laser-photon interaction, normalised to a given laser spot size ($w_0$) as a function of the classical non-linearity parameter $\xi$ for ten days of data acquisition ($N_{day})$, assuming a backround rate of 0.01 ($N_{BG}$) per interaction.  Predictions for full QED and purely perturbative QED are added. Error bars are dominated by statistical uncertainties. The correlated uncertainty of about 40\%, as highlighted in the lower panel, arises from a 5\% uncertainty on the laser intensity. Figure recreated from~\cite{CDR}.}
 \label{fig:gamma-laser}
\end{figure}

In Fig.~\ref{fig:gamma-laser}, the number of expected positrons as a function of $\xi$ is shown for laser-photon interactions.
For $\xi \gtrsim 1$, the difference between the predictions of non-perturbative and perturbative QED with respect to the number of created positrons is expected to become measurable.

Positron tracks are reconstructed from energy deposits on four consecutive pixel detector layers. There is no electromagnetic field within the detector system.
A brief overview of the underlying physical processes and the experimental setup is given in sections 1 and 2 of Ref. \cite{ACAT}.

\section{Data sets and selection}
In this study, simulated data is used. Signal interactions at the interaction point (IP) are generated with PTARMIGAN~\cite{ptarmigan}, a custom Monte Carlo event generator. Positrons stemming from PTARMIGAN are propagated through a dipole magnet and through the positron tracking system using a simplified simulation. In the simplified simulation, parameters such as position and resolution of detector layers, as well as the scattering processes can be tuned to explore the impact on the reconstruction algorithm. In this study, a simplified detector geometry is used, namely four non-overlapping layers.

The used simulated data predicts future measurements of the phase-0 of LUXE for the e-laser setup for $\xi\in$ \{4, 5, 7\}, which corresponds to between 2,000 and 70,000 expected positrons. The 500 particles per laser-beam interaction closest to the beamline are considered for the track reconstruction task in order to equalise the size of all used data sets. The track density increases with $\xi$, thus increasing the complexity of the track reconstruction task.

Our starting point for track reconstruction is either doublets or triplets of energy deposits (hits). Doublets are two connected hits from strictly consecutive layers, while triplets are composed of two doublets with one shared hit. Triplets consist of three hits from strictly consecutive layers. With respect to the beamline, an angle-based pre-selection procedure is applied to the doublets based on the experiment geometry. Triplets are formed if the angles between two doublets with one shared hit do not exceed a threshold value. In this procedure, the combinatorial triplet candidates are reduced without lowering the efficiency.

\section{Methodology}

\subsection*{Classical tracking}
As a benchmark, a classical tracking approach with a combinatorial Kalman filter (CKF) technique is used. For this, a reconstruction software based on A Common Tracking Software (ACTS) toolkit~\cite{ACTS} is employed. Triplets are used as seeds to find an initial estimate of the track parameters. This initial estimate is updated by scanning for matching hit candidates, and the measurement search is performed at the same time. Eventually, after the track finding and fitting is completed, an ambiguity-solving step is applied to remove tracks with shared hits. 

\subsection*{Graph neural network}
Another method explored in this work is the use of a graph neural network (GNN)~\cite{GNN1, GNN2}. Here, hits are represented as nodes. Edges are connections between these nodes, forming doublet-like structures, called segments, and are only kept if they satisfy the pre-selection criteria. The GNN consists of an alternating EdgeNetwork and NodeNetwork and is trained on weighted examples to optimise the edge connections, thus learning which segments should be chosen to be a part of track candidates. We note that there is also a hybrid quantum-classical version of the GNN-based tracking~\cite{GNN3}, which is not examined here but could be studied in future work.

\subsection*{Quantum algorithm}
In the quantum computing part, only triplet-level information is used for the pattern recognition task.
Triplets are encoded as binary variables. A decision is then made on which triplets to keep or to discard in the subsequent track reconstruction process. An objective function is defined, called Quadratic Unconstrained Binary Optimisation (QUBO), as proposed in Ref.~\cite{QUBO}. The goal is to minimise this objective function, 
\beq{O =  \sum_i^N \sum_{j<i} b_{ij} T_i T_j +\sum_{i=1}^{N} a_i T_i, \hspace{2cm} T_i,T_j \in \{0, 1\} },
\eeq{QUBO} 
where $T_i$ and $T_j$ represent triplets at the positions $i$ and $j$ of a possible solution vector, and $a_i$ and $b_{ij}$ are the QUBO coefficients. 

In Eq.~(\ref{QUBO}), the quadratic term describes the relation between triplets, which is quantified by the parameter $b_{ij}$. This parameter has a negative value if the triplets form a track candidate, a positive value if they are in conflict, and a zero value if they do not share a hit. The parameter $a_i$ evaluates a triplet based on the angle between the two doublets that make up the triplet. In this work, we discard the linear term and focus entirely on the quadratic term.

We map the QUBO onto an Ising Hamiltonian, in order to solve it using a simulated quantum device. Finding the ground state of the Ising Hamiltonian is equivalent to minimising the QUBO and thus to finding an optimal solution to the track reconstruction task. The Ising Hamiltonian,

\beq{\mathcal{H}= -\sum_{n=1}^N\sum_{m<n} \bar{b}_{nm}\sigma_n^x \sigma_{m}^x-\sum_{n=1}^N\bar{a}_n \sigma_n^x},
\eeq{Ising}
is solved using the Variational Quantum Eigensolver (VQE), a hybrid quantum-classical algorithm. For this task, the Qiskit~\cite{Qiskit} toolkit is employed. As a benchmark for the VQE, an analytical solution can be obtained by using the Numpy Eigensolver for small QUBOs. We use an ideal, noiseless simulation for VQE. For the optimiser, the Nakanishi-Fujii-Todo (NFT)~\cite{NFT} algorithm is selected. 

\begin{figure}[!htb]
\centering
 \includegraphics[width=0.9\linewidth]{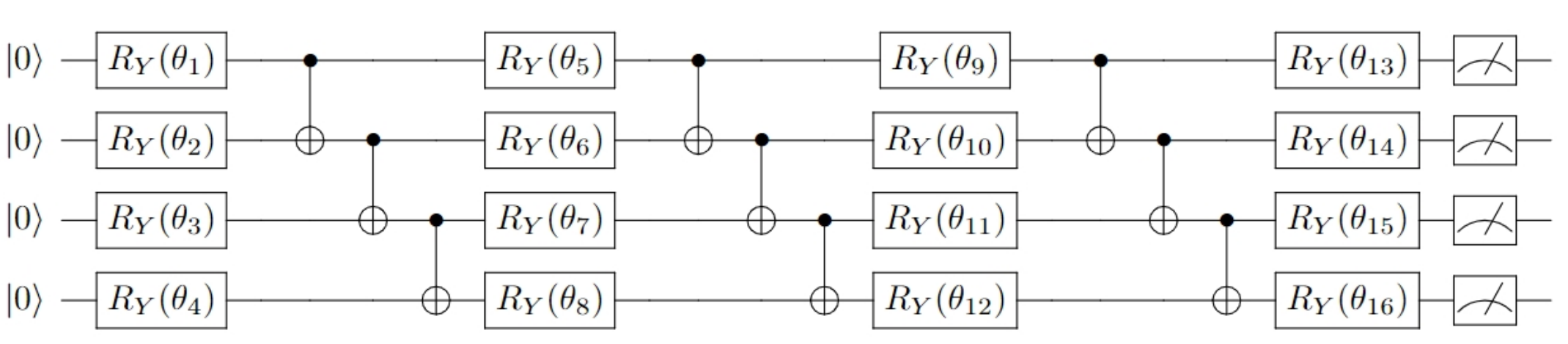}
 \caption{Variational quantum circuit layout. The \textit{TwoLocal} ansatz is used with three repetitions of $R_Y$ and CNOT gates and an additional final rotation layer. For simplicity, only a four qubit system is shown.}
 \label{fig:ansatz circuit}
\end{figure}

We have improved the hyperparameters of the VQE to boost the performance compared to our previous results~\cite{ACAT}. After comparing the results of NFT and Constrained Optimisation by Linear Approximation (COBYLA), we decided to use NFT because it performs better. The quantum circuit following Qiskit's \textit{TwoLocal} ansatz scheme is altered from a circular entanglement scheme to a linear entanglement scheme, and the circuit depth is increased to three. The quantum circuit is shown in Fig.~\ref{fig:ansatz circuit}.

\begin{figure}[!htb]
\centering
 \includegraphics[width=1.0\linewidth]{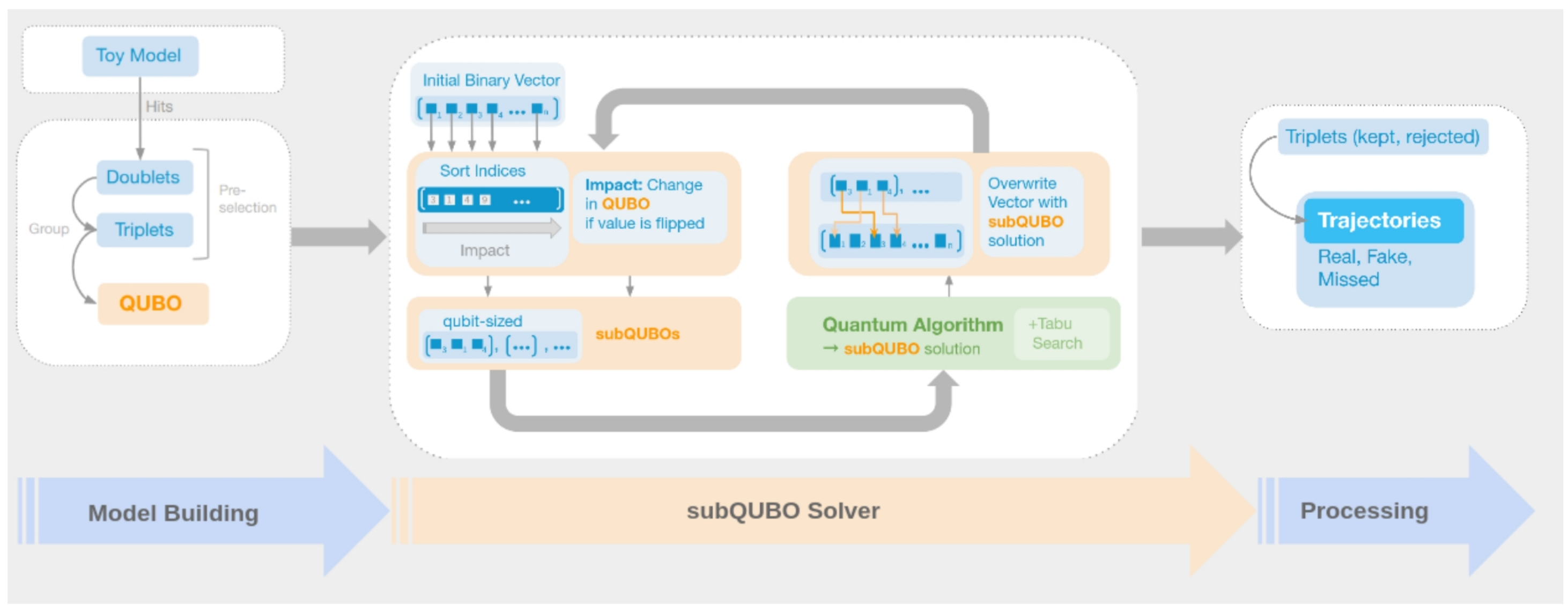}
 \caption{Sketch of the QUBO solving approach with focus on the sub-QUBO routine~\cite{ACAT}. }
 \label{fig:sub-QUBO}
\end{figure}

To solve the QUBO, an initial guess of the solution is made, in the form of a string representation of the set of triplets, assuming values \{$0,1$\}. For solving the QUBO in one step, the number of available qubits for the computation has to be at least the same as the number of triplets participating in the QUBO. For 500 tracks we get approximately 1500 to 3000 triplets, depending on $\xi$ and on the parameters of our pre-selection.
Since sizes of quantum devices of this magnitude are not available and simulating huge devices is computationally infeasible, the problem has to be broken down into smaller parts, called sub-QUBOs, which are solved sequentially in each iteration. A sub-QUBO size of seven is chosen. The order of triplets used in the sub-QUBO process is determined by their impact on the Hamiltonian energy. This impact is defined as the absolute difference between the original objective function and the objective function after flipping the binary representation of the triplet in the QUBO, i.e., for each triplet, the state of the triplet is flipped and the objective function is recomputed. The triplets are then ordered by their impact, from lowest to highest, and partitioned into size-7 sub-QUBOs. All the resulting sub-QUBOs are then solved sequentially. These steps are repeated until a stable solution is found. A sketch of the QUBO solving process with a focus on the sub-QUBO routine is shown in Fig.~\ref{fig:sub-QUBO}.

\section{Results}

Comparing the performance of the different track reconstruction methods is done on track level. Results are averaged over 10 bunch crossings for each $\xi$. Efficiency and fake rate are used as metrics.
A track is defined as a set of four hits of consecutive layers, which is either obtained by combining doublets and triplets into quadruplets or is found directly with the classical CKF-based tracking method. A matched track stems from exactly one particle, fake tracks from multiple particles. 

\begin{figure}[!htb]
\centering
 \includegraphics[width=1.0\linewidth]{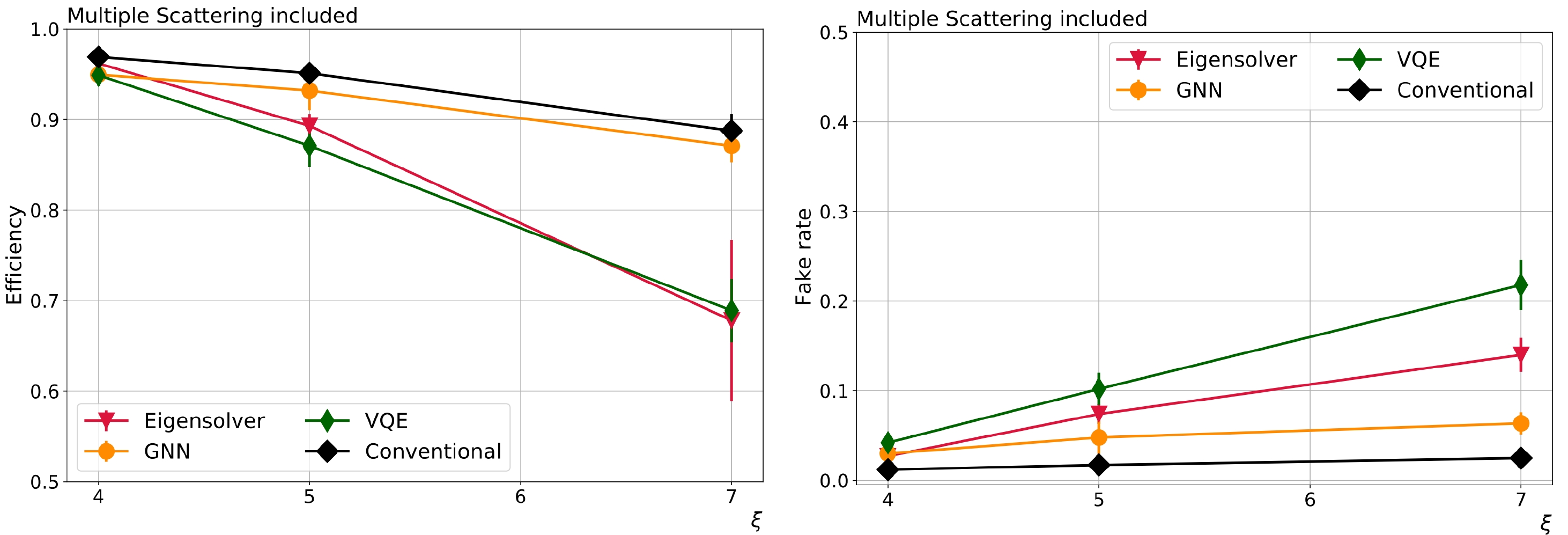}
 \caption{Efficiency (left) and fake rate (right) as a function of $\xi$. }
 \label{fig:500}
\end{figure}

The efficiency and fake rate are defined as

\beq{\textrm{Efficiency} = \frac{N_{\rm tracks}^{\rm matched}}{N_{\rm tracks}^{\rm generated}} \qquad \textrm{and}\qquad
  \textrm{Fake rate} = \frac{N_{\rm tracks}^{\rm fake}}{N_{\rm tracks}^{\rm reconstructed}}\, . }
\eeq{Efficiency_fake_rate}
In Fig.~\ref{fig:500}, the efficiency and fake rate for 500 tracks are displayed as a function of the classical non-linearity parameter $\xi$. Conventional CKF-based tracking is used as a benchmark for efficiency and fake rate, and is compared to GNN-based tracking and VQE. The Eigensolver results are added as a benchmark for the VQE approach for a sub-QUBO size of seven.

\begin{figure}[!htb]
\centering
 \includegraphics[width=1.0\linewidth]{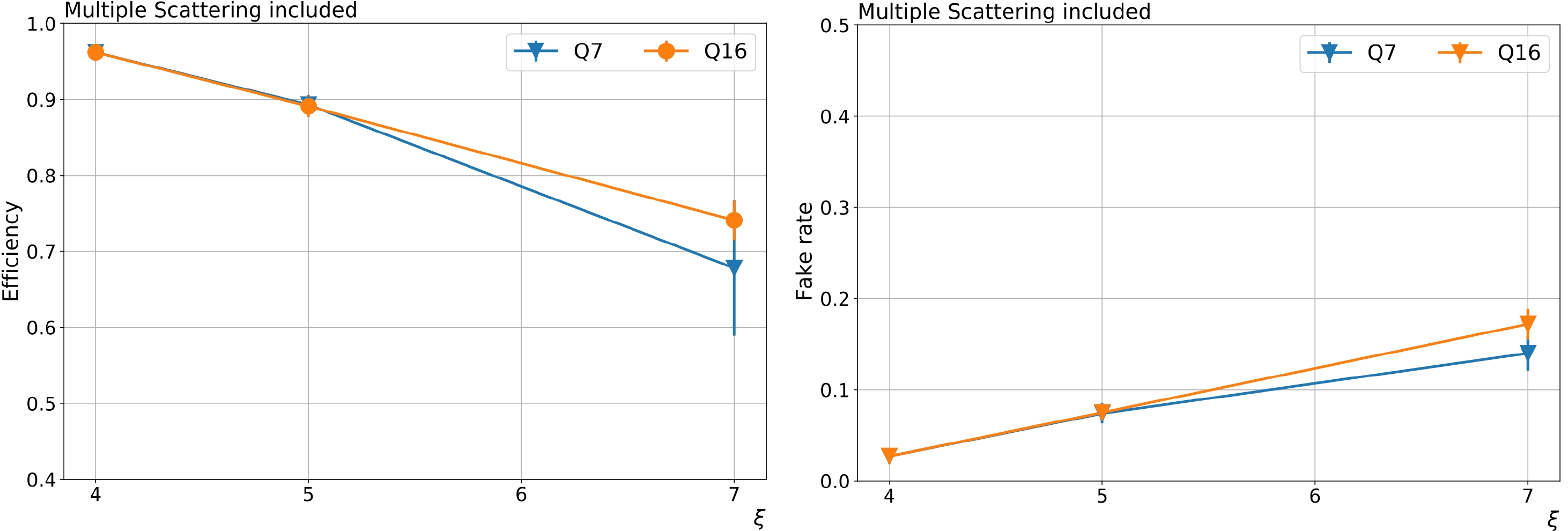}
 \caption{Efficiency (left) and fake rate (right) for 500 tracks as a function of $\xi$ for sub-QUBO sizes 7 (Q7) and 16 (Q16).}
 \label{fig:1000}
\end{figure}

The CKF-based tracking efficiency decreases with $\xi$ but is still performant at the highest shown track density at $\xi=7$.
The GNN-based tracking shows similar, but slightly worse preformance than the CKF-based tracking. The VQE and Eigensolver deteriorate strongly at high $\xi$ values while being comparable with CKF-based tracking and GNN at $\xi=4$. While the performance of the GNN is known to be improved by using more training examples, the VQE performance is likely limited by the set of the QUBO parameters $a_i$ and $b_{ij}$ as well as by the size of the sub-QUBOs, therefore both approaches can be further optimised. To investigate the impact of the sub-QUBO size on the performance, only the Eigensolver is used, because simulating VQE for a higher number of qubits is computationally intensive. 
In Fig.~\ref{fig:1000}, the results for both sub-QUBO sizes is shown. Increasing the size of the sub-QUBOs from 7 to 16 does not yield an improvement in efficiency. However, the statistical uncertainty for sub-QUBO size of 16 is much smaller. 

\begin{figure}[!htb]
\centering
 \includegraphics[width=1.0\linewidth]{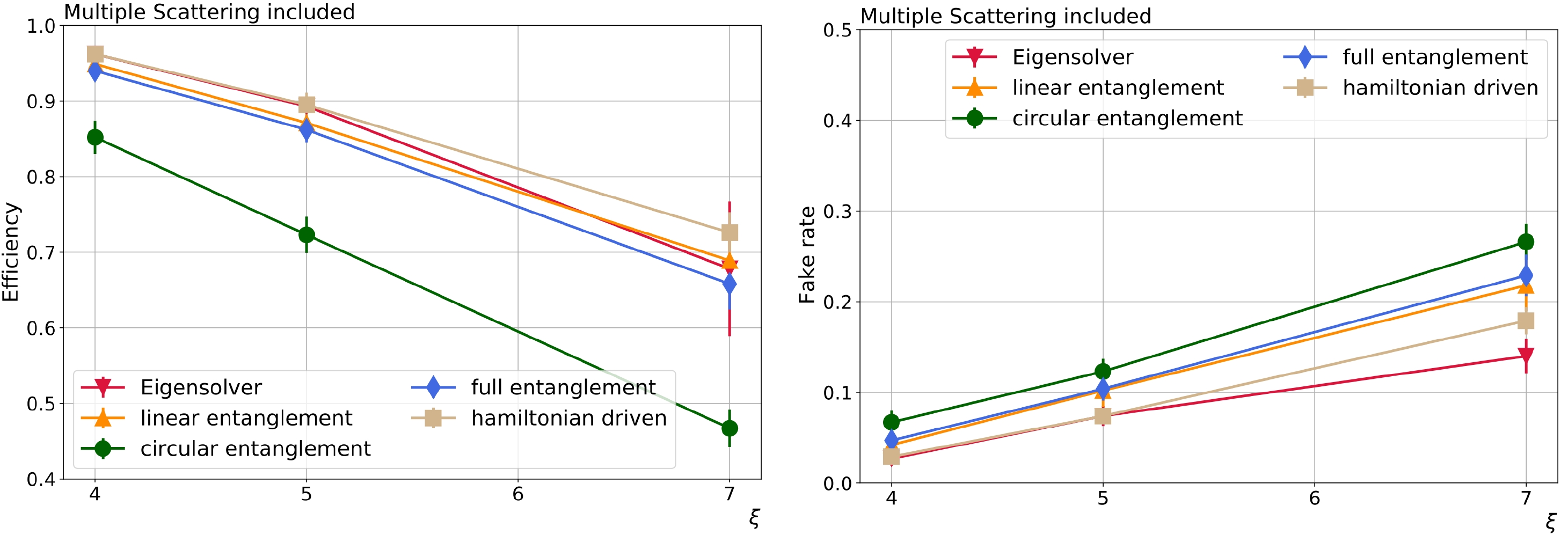}
 \caption{Efficiency (left) and fake rate (right) as a function of $\xi$. The Eigensolver result is the upper limit of what can be achieved by employing VQE and using the sub-QUBO subroutine approach.}
 \label{fig:ansatz_choices}
\end{figure}

Using customised entanglement structures may increase the performance of the VQE on the sub-QUBO level. Four different entanglement structures are compared in Fig.~\ref{fig:ansatz_choices}. Linear entanglement is shown in Fig.~\ref{fig:ansatz circuit}. circular entanglement, as used in our previous study in Ref.~\cite{ACAT}, has an additional CNOT gate, connecting the last to the first qubit. Full entanglement refers to each qubit being entangled with every other qubit. A Hamiltonian driven approach is used if qubits are only connected via a CNOT gate, i.e., if they represent triplets, which actually have a shared hit and thus an immediate connection. The Eigensolver result is an upper limit for the performance of the VQE based solving. The Hamiltonian driven entanglement shows a performance similar to the Eigensolver at low $\xi$, and on average, exceeds the Eigensolver performance  at $\xi = 7$. This is due to the large uncertainty of the Eigensolver result for $\xi = 7$ in combination with a better random initial guess, since it is not possible for the VQE approach to achieve better efficiencies than the Eigensolver. Linear and full entanglement perform slightly worse, whereas circular entanglement shows significantly lower efficiency and higher fake rate. 

\section{Conclusions}

In this work, we updated our previous results~\cite{ACAT} of studying a hybrid quantum-classical algorithm for track reconstruction in the LUXE experiment, as well as a GNN-based tracking approach. As a benchmark, conventional CKF-based tracking is used. With the current version, the quantum approach is less performant than GNN-based and conventional tracking, but we identified several clues for optimisation on the quantum side as well as on the classical side, which will be further investigated in the future. Especially the optimisation of the sub-QUBO routine is a strong candidate for a major improvement.

\vspace{2cm}
\Acknowledgements
{The work by B.H., A.K., F.M., D.S. and Y.Y. was in part funded by the Helmholtz Association - “Innopool Project LUXE-QED”. A.C, K.J. and C.T. are supported in part by the Helmholtz Association - “Innopool Project Variational Quantum Computer Simulations (VQCS)”. L.F.\ is supported by the U.S.\ Department of Energy, Office of Science, National Quantum Information Science Research Centers, Co-design Center for Quantum Advantage (C$^2$QA) under contract number DE-SC0012704, by the DOE QuantiSED Consortium under subcontract number 675352, by the National Science Foundation under Cooperative Agreement PHY-2019786 (The NSF AI Institute for Artificial Intelligence and Fundamental Interactions, http://iaifi.org/), and by the U.S.\ Department of Energy, Office of Science, Office of Nuclear Physics under grant contract numbers DE-SC0011090 and DE-SC0021006. S.K.\ acknowledges financial support from the Cyprus Research and Innovation Foundation under project ``Future-proofing Scientific Applications for the Supercomputers of Tomorrow (FAST)'', contract no.\ COMPLEMENTARY/0916/0048, and ``Quantum Computing for Lattice Gauge Theories (QC4LGT)'', contract no.\ EXCELLENCE/0421/0019. This work has benefited from computing services provided by the German National Analysis Facility (NAF). This work is supported with funds from the Ministry of Science, Research and Culture of the State of Brandenburg within the Centre for Quantum Technologies and Applications (CQTA).}

\end{document}